\documentclass[epj]{svjour}
\usepackage{epsfig}
\newcommand{\beq}{\begin{equation}}
\newcommand{\eeq}{\end{equation}}
\newcommand{\bdis}{\begin{displaymath}}
\newcommand{\edis}{\end{displaymath}}
\newcommand{\bea}{\begin{eqnarray}}
\newcommand{\eea}{\end{eqnarray}}
\newcommand{\barr}{\begin{array}}
\newcommand{\earr}{\end{array}}

\begin{document}

\title{Planar cracks in the fuse model}

\author{Stefano Zapperi\inst{1,2} \and  Hans J. Herrmann\inst{2}
\and St\'ephane Roux\inst{3}}

\institute{
INFM sezione di Roma 1, Universit\`a "La Sapienza", 
P.le A. Moro 2 00185 Roma, Italy.\and
PMMH-ESPCI, 10 Rue Vauquelin, 75231 Paris cedex 05, France\and
Laboratoire Surface du Verre et Interfaces
Unit\'e Mixte de Recherche CNRS/Saint-Gobain,
39 Quai Lucien Lefranc, BP. 135,
F-93303 Aubervilliers Cedex, France.}

\abstract{
We simulate the propagation of a planar crack in a quasi-two dimensional
fuse model, confining the crack between two horizontal plates.
We investigate the effect on the roughness of microcrack nucleation 
ahead of the main crack and study the structure of the damage zone.
The two dimensional geometry introduces a characteristic length
in the problem, limiting the crack roughness. The damage ahead of the
crack does not appear to change the scaling properties of the
model, which are well described by gradient percolation.
}

\date{\today}
\maketitle

\section{Introduction}

Understanding the mechanisms of crack propagation is an important 
issue in mechanics, with potential application to geophysics and 
material science. Experiments have shown that in several materials
under different loading conditions, the crack front tends to 
roughen and can often be described by self-affine scaling \cite{man}.
In particular, the {\it out of plane} roughness exponent is found
displays universal values for a wide variety of materials \cite{bouch}. 
Interesting experiments have been recently performed on 
PMMA and the {\it in plane} roughness of a planar crack
was observed to scale with an exponent $\zeta=0.63\pm0.03$ \cite{schmit1}.

While the experimental characterization of crack roughness
is quite advanced and the numerical results very accurate,
theoretical understanding and numerical models are still
unsatisfactory. The simplest theoretical approach to the
problem identifies the crack front with a deformable line
pushed by the external stress through a random toughness
landscape.  The deviations of the crack front from 
a flat line are opposed by the elastic stress field, through
the stress intensity factor \cite{gao}. In certain conditions, the
problem can be directly related to models and theories of interface
depinning in random media and the roughness exponent computed
by numerical simulations and renormalization group calculations 
\cite{natt,nf}.
Unfortunately, the agreement within this theoretical approach
and experiments are quite poor. For the out of plane roughness
the theory predicts only a logarithmic roughness in mode I \cite{ram1},
while the experimental results give $\zeta_{\perp}=0.5$
at small length scale and $\zeta_{\perp}=0.8$ at larger 
length scales \cite{bouch}. 
For planar cracks, simulations predict $\zeta=0.35$ \cite{schmit2}
and RG gives $\zeta=1/3$ \cite{ram2}, both quite far from the experimental
result. The inclusion of more details in the model, 
such as elastodynamic effects, does not lead to better results 
\cite{ram2}.

A different approach to crack propagation in disordered
media, considers the problem from the point of view of 
lattice models \cite{hr}. The elastic medium is replaced by a network
of bonds obeying discretized equations until the stress 
reached a failure threshold. The disorder in the medium can be
simulated by a distribution of thresholds or by bond dilution.
Models of this kind have been widely used in the past to 
investigate several features of fracture of disordered media,
such as the failure stress \cite{fuse,duxbury,fuse2,fuse3}, 
fractal properties \cite{hr,th} and avalanches \cite{hh,th,zrsv,gcalda,zvs}.
The out of plane roughness exponent has been simulated in
two dimension, resulting in $\zeta_{\perp}\simeq 0.7$ \cite{HHR,CCG}, and 
three dimensions where 
$\zeta_{\perp}\simeq 0.4-0.5$ \cite{RSAD,BH-98,pcp}.
The last result is in good agreement with experimental 
results, if we identify the small length scales with the
quasistatic regime used in simulations. The advantage
of lattice models over interface models is that the former
allow for nucleation of microcracks ahead of the main 
crack. While it is well known experimentally that
microcracks do nucleate, their effect on the roughness
exponent has never been studied. 

In this paper we present numerical simulations of a
planar crack using the random fuse model \cite{fuse}. We employ a
quasi two-dimensional geometry, considering two horizontal
plates separated by a network of vertical bonds.
A similar setup was used in a spring model \cite{nak}, but
the roughness was studied only in the high velocity regime
for crack motion. The experiments of Ref. \cite{schmit1} where instead
performed at low velocity so that a quasistatic model 
seems more appropriate. 

We find that the two dimensional geometry introduces
a characteristic length limiting the crack roughness.
In addition, crack nucleation does not appear to change in a
qualitative way the behavior of the system. For length scales
smaller than the characteristic length, the crack is not
self affine, but possibly self-similar. We study the 
damage zone close to the crack and find that several of its
features can be described by gradient percolation \cite{gradper}.

\section{Model}

\begin{figure}[htb]

\centerline{
        \epsfxsize=8.0cm
        \epsfbox{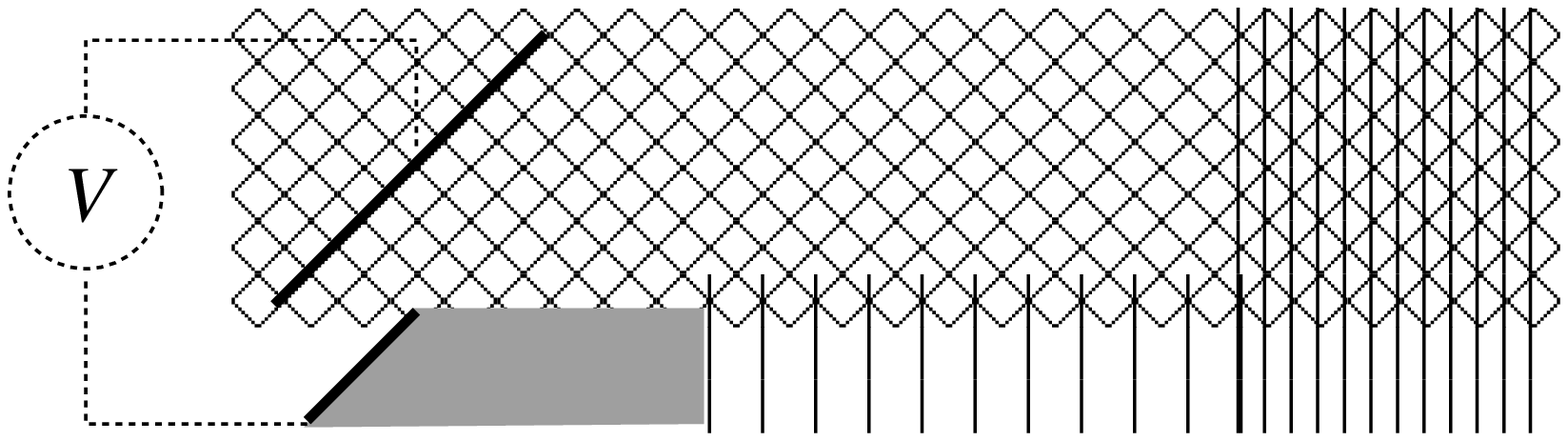}
        \vspace*{0.5cm}
        }
\caption{The geometry of the model. The horizontal bonds have
unitary conductivity while the vertical bond have conductivity
$\sigma$. A planar crack is present at the center of the system
and a voltage drop is applied at the boundaries.
}
\label{fig:0}
\end{figure}

In the random fuse model, each bond of the lattice represents
a fuse, that will burn when its current overcome a threshold 
\cite{fuse,duxbury,fuse2,fuse3}.
The currents flowing in the lattice
are obtained solving the Kirchhoff equation with appropriate
boundary conditions \cite{nota_kir}. 
In this paper, we consider two horizontal
tilted square lattices of resistors 
connected by vertical fuses (see Fig.~\ref{fig:0}). The conductivity
of the horizontal resistors is chosen to be unity, while the vertical
fuses have conductivity $\sigma$. A voltage drop $\Delta V$ is
imposed between the first horizontal rows of the plates. To
simulate the propagation of a planar crack, we allow for failures
of vertical bonds only and assign to each of them a random threshold
$j^c_i$, uniformly distributed in the interval $[1:2]$. 
When the current in a bond $i$ overcomes the random threshold,
the bond is removed from the lattice and the currents in the lattice
are recomputed, until all the currents are below the threshold.
The voltage drop is thus increased until the weakest bond 
reaches the threshold. 

The quasistatic dynamics we are using should correspond
to the small constant displacement rate at the boundary of
the crack used in experiments \cite{schmit1}.  
In order to avoid spurious boundary effect, we start 
with a preexisting crack occupying the first half of the lattice
(see  Fig.~\ref{fig:0})  and employ periodic boundary conditions 
in the direction parallel to the crack. In addition, once an entire 
row of fuses has failed, we shift the lattice backwards 
one step in the direction perpendicular
to the crack, to keep the crack always in the middle
of the lattice. 

Before discussing the numerical results, we present some
analytical considerations which will guide the simulations.

\section{Characteristic length}

Here we investigate the model introduced in the preceding section
in some particular configuration. We first analyze the case of
a perfectly straight planar crack and study the current decay 
in front of it. In this condition, the system is symmetric
in the direction parallel to the crack and we can thus reduce it
to one dimension. 

We consider an infinite  ladder composed of vertical bonds of
resistance $r\equiv 1/\sigma$ connected by unitary horizontal 
resistances. 
Since the ladder is infinite, we can add one additional step without
changing the end to end resistance $R$:
\begin{equation}
R=2+1/(1/r+1/R)=2+{rR \over r+R}.\label{eq:R}
\end{equation}
Solving Eq.~(\ref{eq:R}) we obtain the total resistance 
\begin{equation}
R=\sqrt{1+2r}+1.
\end{equation}
The fraction of current $j$ flowing through the first ladder step is such
that $rj=R(1-j)$, which implies
\begin{equation}
j={R\over r+R}={\sqrt{1+2r}+1\over r+1+\sqrt{1+2r}}
={\sqrt{1+2r}-1\over r}.
\end{equation}
The current flowing in the second ladder step is then $(1-j)j$ and 
similarly in the $n$th step it is given by
$j_n=(1-j)^{n-1} j$, thus scaling as 
$j_n\propto \exp(-n/\xi)$ where 
\begin{equation}
\xi\equiv {-1\over\log(1-j)}={-1\over\log(1-\sigma(1-\sqrt{1-2/\sigma}))}.
\end{equation}

Thus the current in front of the crack decays exponentially 
with a characteristic length
$\xi \simeq 1/\sqrt{2\sigma}$, for $\sigma\ll 1$.
A similar result could have been anticipated from the structure 
of the Kirchhoff equations reading as
\begin{equation}
\sum_{nn} (V_{i+nn}-V_i)+\sigma (V^\prime_i-V_i)=0
\label{eq:vdis}
\end{equation}
where the sum runs over the nearest neighbors of node $i$
and $V^\prime_i$ is the voltage of the corresponding 
node in the opposite plate. Due to symmetry we can chose
$V_i^\prime = -V_i$ and solve the equations only for one
of the plates. Eq.~(\ref{eq:vdis}) represents a discretization
of a Laplace equation with a ``mass term'' 
\begin{equation}
\nabla^2 V - \xi^{-2} V  = 0,
\label{eq:vcont}
\end{equation}
where $\xi=1/\sqrt{2\sigma}$.

The continuum limit can be used to understand how current is
transfered after a single failure. We define $G(x-x^\prime)$ as the 
difference in the currents in $x$ before and after a bond
in $x^\prime$ has failed. The function $G$ is analogous of 
the ``stress Green function'' used in interface models
for cracks propagation \cite{ram1,schmit2,ram2}. 
In fact, the equation of motion in these models is written as
\begin{equation}
\frac{\partial h(x,t)}{\partial t} = F + \int dy G(x-y)(h(y,t)-h(x,t))
+\eta(x,h),
\end{equation}
where $h$ indicate the position of the crack, $F$ is proportional
to the external stress, $\eta$ to the random toughness of the
material and for a planar crack in three dimensions $G(x)\sim 1/|x|^2$.
A renormalization group analysis shows that the roughness of the 
interface crucially depends on the decay of $G$ \cite{natt,nf}. If $G$ decays 
slower than $|x|^{-1}$, for $x\to \infty$ the interface is not rough
on large length scales.

In order to compute the function $G$, we solve Eq.~(\ref{eq:vcont})
with the appropriate boundary conditions. Note that by definition
$G(x)$ is proportional to the differences in the voltages in $x$
before and after removing a fuse. Since Eq.~(\ref{eq:vcont}) is
linear, the difference of the voltages still satisfies the equation
in all points except $x=0$. This condition can also be expressed
in terms of the current $J\equiv\partial V/\partial y$ \cite{cond},
which should be continuous everywhere apart from $x=0$.

Let's consider a  planar crack along the  $x$ direction and identify
two domains:\hfill\break 1) the domain where fuses are present ($y>0$)
labeled {\bf A} \hfill\break 2) the domain where all fuses
are burnt out ($y<0$) labeled {\bf B}.  Thus the equation 
to solve in domain {\bf A} is 
\begin{equation}
\nabla^2 V=\xi^{-2}V
\end{equation}
and in {\bf B} $\nabla^2 V=0$.

Taking the Fourier transform along $x$, and calling $k$ the conjugate
variable to $x$, we can write in domain {\bf A}
\begin{equation}
\partial^2_y \tilde V=(k^2+\xi^{-2}) \tilde V.
\end{equation}
Integrating the equation, setting $V\to 0$ at infinity, we obtain
\begin{equation}
\tilde V(k,y)=\tilde V(k,0) \exp(-y/\ell)
\end{equation}
where $1/\ell=\sqrt{k^2+\xi^{-2}}$.
A similar calculation allows to obtain $\tilde V(k,y)$
in domain {\bf B}. 

The currents normal to the crack in the two domains are given by 
\begin{equation}
\tilde J_A(k)=-\sqrt{k^2+\xi^{-2}}\tilde V(k,0)
\end{equation}
and
\begin{equation}
\tilde J_B(k)=-|k|\tilde V(k,0),
\end{equation}
where $\tilde V$ is the same for the two domains. 
If one bond is removed at $x=0$ along the interface, the continuity
of the current implies $J_A+J_B\sim\delta(x)$ 
and in Fourier space   
\begin{equation}
\tilde J_A(k)+\tilde J_B(k) \sim 1
\end{equation}
and hence
\begin{equation}
\tilde V \sim {-1\over |k|+\sqrt{k^2+\xi^{-2}}}.
\end{equation}
The Fourier transform of the function $G$ is simply proportional to 
$\tilde V$ and therefore at short 
distances $k\xi\gg 1$, $\tilde V\propto 1/|k|$, or $ G \propto \log(x)$,
while at long distances $k\xi\ll 1$, $G\propto \exp(-r/\xi)$.

\begin{figure}[htb]

\centerline{
        \epsfxsize=8.0cm
        \epsfbox{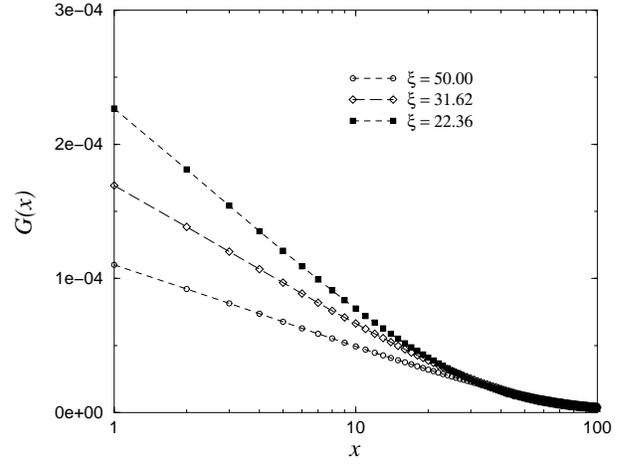}
        \vspace*{0.5cm}
        }
\centerline{
        \epsfxsize=8.0cm
        \epsfbox{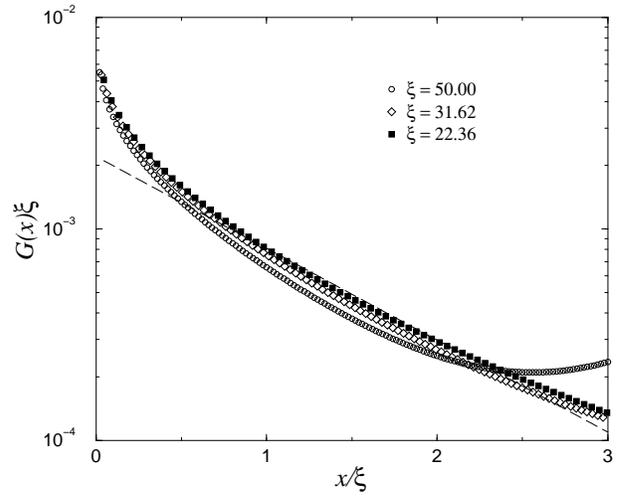}
        \vspace*{0.5cm}
        }
\caption{The current transfer function in log-log plot (top),
to show the logarithmic behavior at short length scales. In the 
linear-log plot (bottom) the function $G$ has been rescaled
with $\xi$ and the collapse is in agreement with the analytical
solution. The deviation from a pure exponential $exp(-x/\xi)$,
plotted as a dashed line, is due 
to the periodic boundary conditions employed in the simulation.}

\label{fig:1}
\end{figure}

We test the asymptotic behavior predicted above by estimating 
$G$ from numerical simulations. The results for a lattice of
size $L=128$ are in good agreement with the analytical predictions
as shown in Fig.~\ref{fig:1}.
It is interesting to remark that the roughness of the crack
is limited by $\xi$, but even in the limit $\xi\to\infty$ we
do not expect a self affine crack, since $G$ decays slower
than $1/|x|$. In the next section, we will show numerically
that damage nucleation does not alter this conclusion.

\section{Crack roughness: simulations}

\begin{figure}[htb]

\centerline{
        \epsfxsize=8.0cm
        \epsfbox{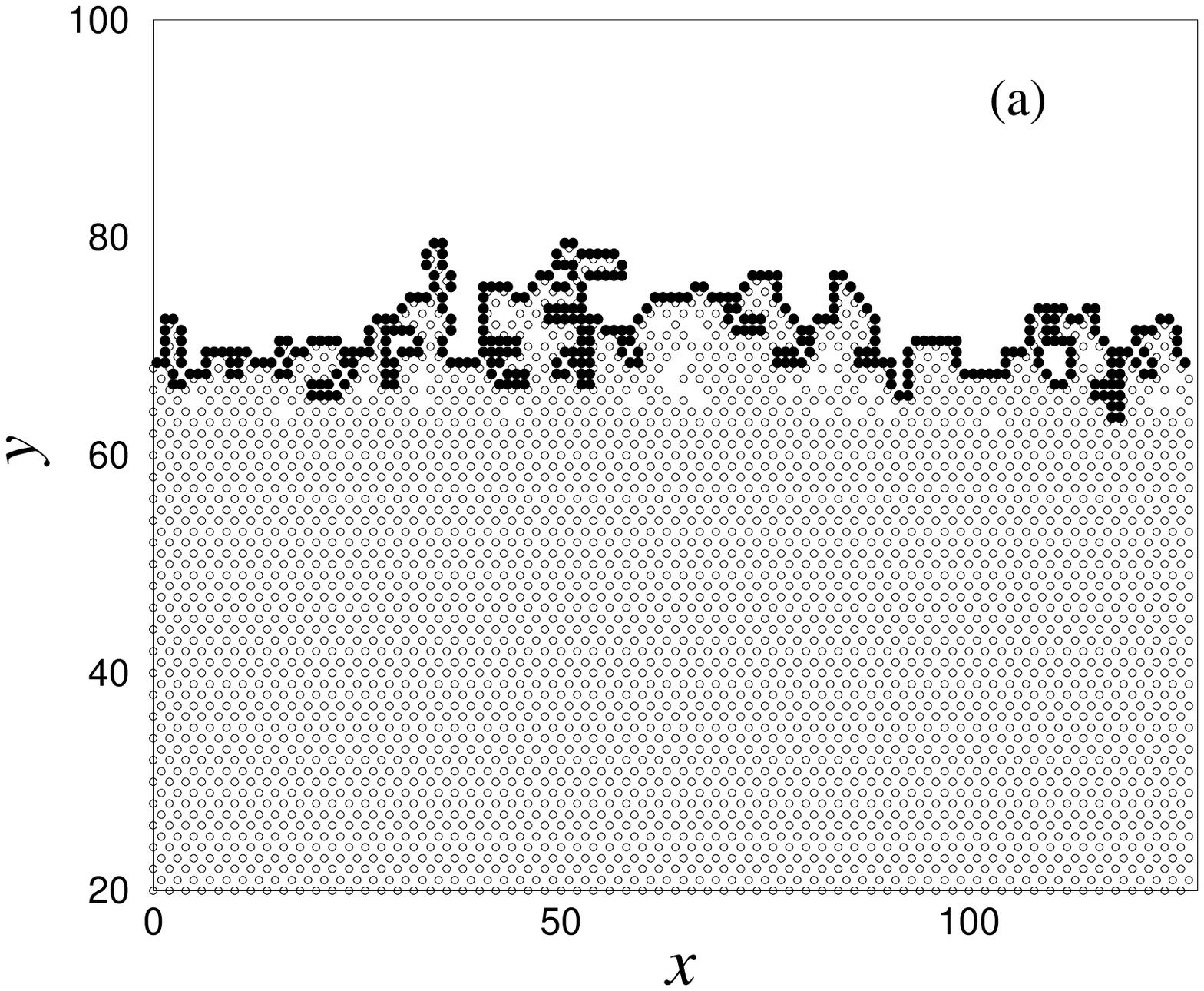}
        \vspace*{0.5cm}
        }
\centerline{
        \epsfxsize=8.0cm
        \epsfbox{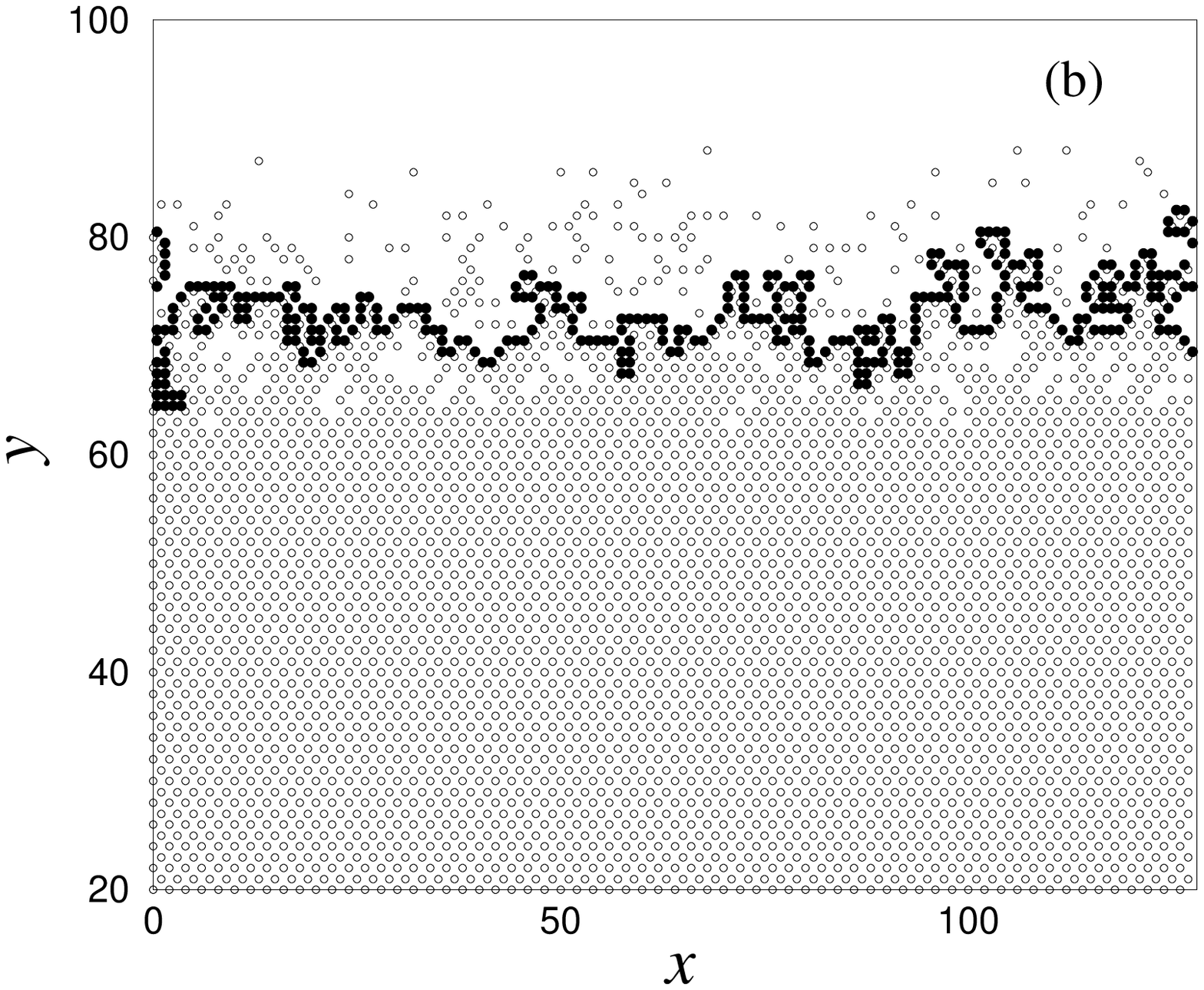}
        \vspace*{0.5cm}
        }
\caption{The crack profiles in the steady state, when only 
a connected crack is allowed to grow (a) and when 
damage is not restricted to occur only close to the crack (b).
The crack interface is shown in dark.}
\label{fig:2}
\end{figure}

\begin{figure}[htb]

\centerline{
        \epsfxsize=8.0cm
        \epsfbox{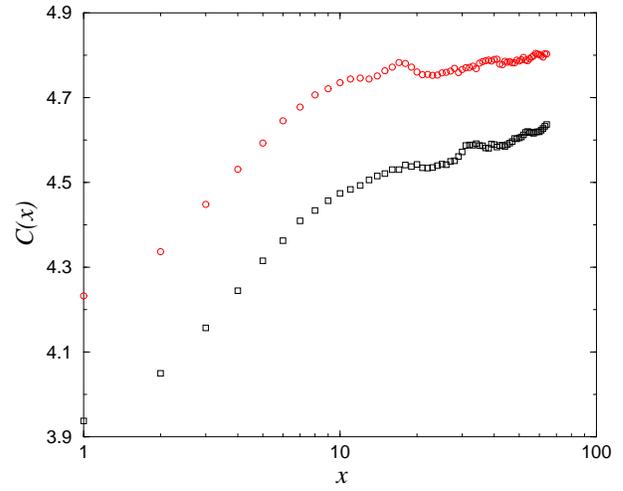}
        \vspace*{0.5cm}
        }
\caption{Comparison between the height correlation function of
a connected crack (bottom curve) and
that of a unrestricted crack (upper curve). 
}
\label{fig:3}
\end{figure}

In order to analyze the effect of crack nucleation ahead
of the main crack, we first simulate the model confining
the ruptures on the crack surface. In this way, our model reduces 
to a connected interface moving in a random medium with an effective
stiffness given by the solution of the Kirchhoff equations.
The results are then compared with simulations of the unrestricted
model, where ruptures can occur everywhere in the lattice. 
In both cases the crack width increases with time up to
a crossover time at which it saturates. In Fig.~\ref{fig:2}
we compare the damage structure in the saturated regime for
the two growth rules. The height-height correlation
function  $C(x)\equiv \langle (h(x)-h(0))^2 \rangle$, where
the average over different realizations of the disorder,
is shown in Fig.~\ref{fig:3}. From these figures it is apparent
that the structure of the crack is similar in the two cases.
The only difference lies in the higher saturation width that
is observed when microcracks are allowed to nucleate ahead
of the main crack.

\begin{figure}[htb]

\centerline{
        \epsfxsize=8.0cm
        \epsfbox{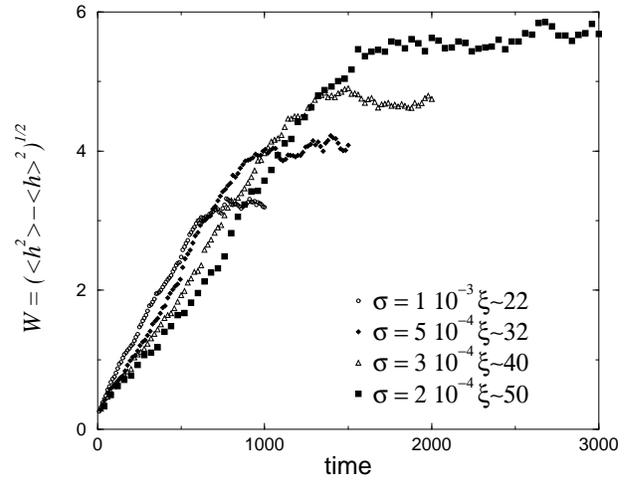}
        \vspace*{0.5cm}
        }
\caption{The global width as a function of time for different
values of $\sigma$ and $\xi=1/\sqrt{2\sigma}$.}
\label{fig:4}
\end{figure}

\begin{figure}[htb]

\centerline{
        \epsfxsize=8.0cm
        \epsfbox{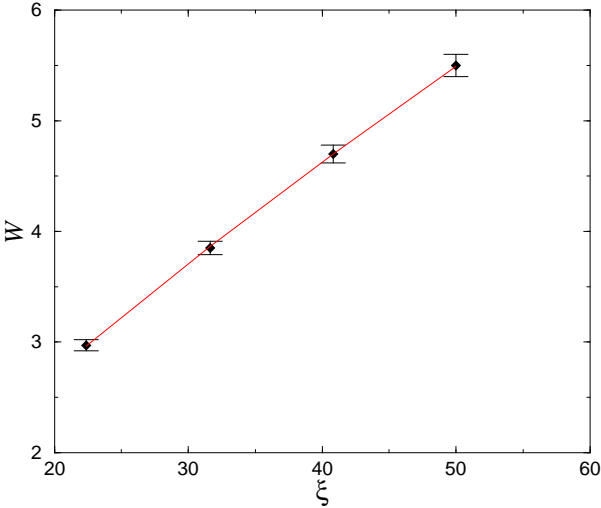}
        \vspace*{0.5cm}
        }
\caption{The saturated global width as a function of $\xi$.}
\label{fig:5}
\end{figure}

Next, we analyze the behavior of the crack as a function
of $\sigma$ which should set the value of the characteristic
length to $\xi \simeq 1/\sqrt{2\sigma}$. In this study we 
restrict our attention to the general model with crack
nucleation. We compute the global
width $W\equiv (\langle h^2\rangle- \langle h\rangle^2)^{1/2}$, averaging
over several realization of the disorder (typically 10),
as a function of time for different values of $\xi$. Fig.~\ref{fig:4}
shows that $W$ increases linearly in time until saturation.
The global width in the saturated regime scales as
$\xi^\zeta$, with $\zeta\simeq 0.75$, 
as shown in Fig.~\ref{fig:5}. Due to the limited
scaling range, we could not obtain a more reliable estimate
of the exponent value.

\section{Mean-field approach}

The long-range nature of the Green function suggests that a mean-field
approach could be suitable.  We outline here the spirit of such an
approach, through the determination of the density of burnt fuses ahead of
the crack front.  First, we note that  
the mean profile is expected to be translational
invariant along the $x$ axis and thus 
the problem reduces to a one dimensional geometry.  
As argued earlier, the tension 
$V(y)$ should obey the following differential equation, in the continuum
limit:
\begin{equation}
{\partial^2 V\over\partial y^2}=2\sigma(y) V(y)
\end{equation}
where $\sigma(y)$ is the ($x$-averaged) conductivity at position $y$.  The
latter can be written as $(1-D(y)) \sigma_0$ where $D(y)$ is the ``damage'',
i.e. fraction of burnt fuses.  This fraction is a known function of the
current density in the mean-field approach.  Namely, the vertical current going
through intact fuses is $j(y)=2\sigma_0 V(y)=V(y)/\xi^2$. 
It is remarkable that 
$D(y)$ drops out of this equation:  the damage reduces the current density
flowing at a given position by a factor $(1-D(y))$ as compared to the intact
state,  but the same current density flows through a reduced number of
intact fuses, and is thus multiplied by $1/(1-D(y))$. In
conclusions the two factors cancel out.

\begin{figure}[htb]

\centerline{
        \epsfxsize=8.0cm
        \epsfbox{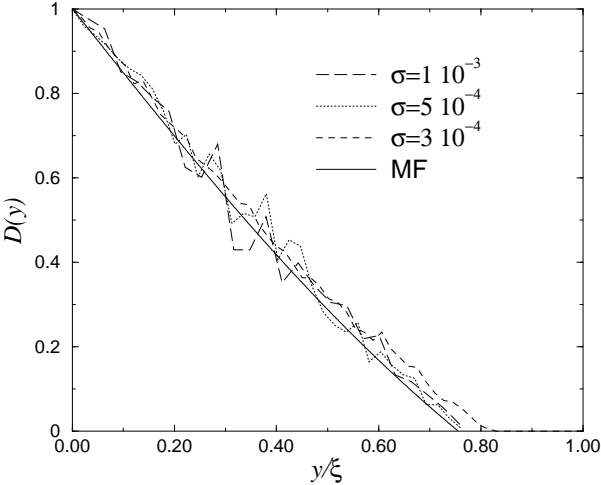}
        \vspace*{0.5cm}
        }
\caption{The average concentration $D(y)$ of burnt fuses as
a function of the reduced distance $y/\xi$ from the crack
for different values of $\sigma$. The numerical results
are in excellent agreement with the mean-field solution.}
\label{fig:6}
\end{figure}

The proportion of fuses which may support the current without burning is
given by the cumulative distribution of threshold currents:
$P(j)=\int_j^\infty p(j') dj'$, which implies $D(y)=1-P(j(y))$.  
The voltage profile along the $y$ axis is thus given by 
\begin{equation}
\xi^2{\partial^2 V\over\partial y^2}= P(V(y)/\xi^2) V(y)
\end{equation}
This equation can be rewritten in terms of the 
rescaled coordinate $s=y/\xi$ and current $j=V/\xi^2$
and in our case, since 
$P(j)=2-j$ for $1<j<2$, we obtain
\begin{equation}
j''(s)=j(s)(2-j(s)).
\label{eq:j}
\end{equation}
Notice that Eq.~\ref{eq:j} is valid only for  $1<j<2$, while for
$j<1$ the equation becomes $j''=j$ and for $j>2$ we have $j''=0$.
At infinity, $j<1$ thus the current is given by
$j=e^{-(s-s_0)}$, for $s>s_0$, so that the boundary
condition for Eq.~\ref{eq:j} at $s=s_0$ is $j(s_0)=-j'(s_0)=1$
With these boundary conditions, 
Eq.~(\ref{eq:j}) can not be solved explicitely so we
resort to numerical integration. From the solution of
Eq.~(\ref{eq:j}) we obtain the damage profile and 
compare it to numerical simulations (see Fig.~\ref{fig:6})  
The remarkable agreement between the mean-field solution
and simulations, with a damage profile which is a single function of
$y/\xi$, implies  that the fracture front should be given by
gradient percolation (in fact the gradient is non-linear)
\cite{gradper}.  From this observation we can extract 
the scaling of the front with the gradient $g$ (here
$g\propto 1/\xi$) as $W\propto g^{-\nu/(1+\nu)}\propto \xi^{\nu/(1+\nu)}$
where $\nu$ is the percolation correlation length critical exponent
$\nu=4/3$, or $W\propto \xi^{0.57}$, reasonably consistent with our data.

\section{Conclusions}

In this paper, we have studied the propagation of planar cracks
in the random fuse model. This model allows to investigate 
the effect on the crack front roughness 
of the microcracks nucleating ahead of the main crack. 
The study was restricted to a quasi
two dimensional geometry and could apply to cases in which
the material is very thin in the direction perpendicular to
the crack plane \cite{plate}. 

In two dimensions, the geometry of the lattice induces a characteristic
length $\xi$ limiting the roughness and microcrack nucleation
does not appear to be relevant. In addition, for length scales
smaller than $\xi$ the Green function decays very slowly,
suggesting the validity of a mean-field approach.
We study the problem numerically, computing the scaling
of the crack width with time and $\xi$, and analyze the
damage ahead of the crack. The results suggest an interpretation
in terms of gradient percolation \cite{gradper}, as it is also
indicated by mean-field theory. The limited range
of system sizes accessible to simulations does not allow for
a definite confirmation of these results.

The present analysis does not resolve the issue of the origin
of the value of the roughness exponent for planar cracks
in heterogeneous media. While microcrack nucleation is
irrelevant in the present context, three dimensional simulations
are needed to understand whether this is true in general.
In principle, one could still expect that microcrack nucleation 
in three dimensions would change the exponent of the interface
model ($\zeta=1/3$), but the present results
do not lead to such a conclusion.

\section*{Acknowledgment}
S. Z. acknowledges financial support from EC TMR Research Network
under contract ERBFMRXCT960062. We thank  A. Baldassarri, 
M. Barthelemy and J. R. Rice  and A. Vespignani 
for useful discussions.

\begin{figure}[htb]

\centerline{
        \epsfxsize=8.0cm
        \epsfbox{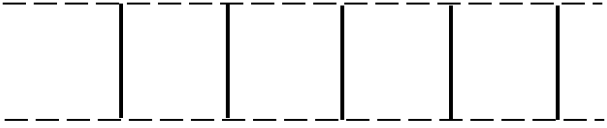}
        \vspace*{0.5cm}
        }
\caption{The one dimensional ladder model used to compute
the characteristic length. The vertical bonds have resistance
$r$ the horizontal have unitary resistance. The end to end resistance
of the ladder is $R$.}
\label{fig:1d}
\end{figure}

\begin{thebibliography}{99}
\bibitem{cond}
        We have dropped the unitary horizontal conductivity.
\bibitem{nota_kir} 
        The equations are solved numerically using 
        a multigrid relaxation method, with precision $10^{-7}$.
\end{thebibliography}
\end{document}